\date{}
\documentclass[12pt]{article}
\usepackage [dvips]{graphicx}
\begin{document}
\title
{Random matrices, random polynomials and Coulomb systems}
\author{\Large P. Leboeuf \hspace{-0.2cm}}
\maketitle
\noindent
\vspace*{-1.5cm}
\begin{center}
{\it Laboratoire de Physique Th\'eorique et Mod\`eles Statistiques 
\\ Universit\'e de Paris-Sud,
B\^at.100 \\ 91405 Orsay Cedex, France}
\end{center}

\vspace*{0.3cm}

\abstract{It is well known that the
joint probability density of the eigenvalues of Gaussian 
ensembles of random matrices may be interpreted as a Coulomb gas.
We review these classical results for hermitian and 
complex random matrices, with special attention devoted to 
electrostatic analogies. We also discuss the joint probability density of the
zeros of polynomials whose coefficients are complex
Gaussian variables. This leads to a new two-dimensional solvable 
gas of interacting particles, with non-trivial interactions 
between particles.}

\vspace*{0.8cm}

\noindent {\sl To appear in the Proceedings of the International Conference on
Strongly Coupled Coulomb Systems, Saint-Malo, 1999 (EDP Sciences, Les Ulis).}

\vspace*{0.6cm}

\section{\hspace*{-0.9cm} .~{\large INTRODUCTION}}

The study of high lying states in the spectrum of atomic nuclei led to the
introduction of a new type of statistical mechanics \cite{wigner,dyson}. Contrary
to the traditional one, this new probabilistic theory does not associate a
weight function to the different configurations a system may take, but rather
introduces a statistical law directly at the level of the Hamiltonian.
Depending on the symmetries of the system, an ensemble of Hamiltonians is
defined with a proper probability distribution function. The
statistical properties of the eigenvalues and eigenvectors of such random
matrices (or averages of physical quantities over such distributions) may then
be compared to a specific system, in which the average is made over some 
energy window.

Motivated by its mathematical tractability, one of the original ensembles
introduced is the so-called Gaussian ensemble of random matrices, defined as
the set of matrices whose elements are Gaussian independent variables.
The probability density of a given $N \times N$ hermitian matrix is defined as
\begin{equation} \label{1}
P(H)={\cal N} \exp\left(-\frac{\beta}{2} {\rm tr} (H^2) \right),
\end{equation} 
where $\cal{N}$ is a normalization constant. A very specific aspect of this
ensemble is its invariance under rotations in Hilbert space. The form of $H$ 
depends on symmetry considerations: its matrix elements are
real for even spin systems with time reversal symmetry, complex when no 
symmetry is present and quaternion for odd spins with time reversal symmetry.
For these three cases, the parameter $\beta$ in Eq.(\ref{1}) takes the value
$1$, $2$ and $4$, respectively.

During the 40 years following its introduction, random matrix theory 
(RMT) has proven to be a pervasive theory (see e.g.\cite{mehta}). Its 
applicability ranges from the description of atomic and
molecular systems \cite{porter}, the behavior of classically chaotic systems
\cite{houches1}, the closely related motion of electrons in a disordered 
potential \cite{houches2}, up to the description of the statistical properties
of the critical zeros of the Riemann zeta function \cite{odlyzko}.
In the second section of this 
paper our aim is to present a brief survey of some elementary results on the 
statistical properties of eigenvalues and eigenvectors associated to 
Eq.(\ref{1}). The most characteristic feature concerning the distribution of 
eigenvalues is the
existence of a strong repulsion acting among them. This repulsion may be seen,
in an electrostatic interpretation, as a two-dimensional Coulomb interaction.
For short distances, this Coulomb interaction is not a
particular feature of the ensemble (\ref{1}) but rather constitutes a very 
basic and general fact of any quantum mechanical system. The general arguments
leading to this result from standard perturbation theory are also presented 
in section 2.

Because of the symmetry of $H$, the eigenvalues lie on the real axis of the
complex plane. A fully two-dimensional Coulomb gas arises when the basic 
hermitian symmetry of
the Hamiltonian is abandoned. We illustrate this in section 3 with the
example of the ensemble of complex matrices introduced by Ginibre 
\cite{ginibre}. We
discuss in some detail the nearest neighbor spacing distribution in the 
complex plane since this quantity has been less studied in two dimensions. 

In the same way the spectrum of random matrices may be associated to
the equilibrium properties of one or two-dimensional Coulomb systems, it is
possible to associate to the eigenstates of random matrices a two-dimensional 
gas of interacting particles (section 4). Although it contains a Coulomb 
repulsion, the full interaction between particles is now more complicated and 
includes many-body terms. 

This reduction of the statistical properties of eigenstates to a
two-dimensional interacting system is done through the Bargmann or coherent
state representation of quantum mechanics. For finite (but large) matrices
like the ones considered here, this representation associates to each
eigenstate a polynomial whose coefficients, according to RMT, are random
independent distributed. The reduction is completed when looking at the
distribution of the roots in the complex plane of these random polynomials.
The joint probability density of the zeros may be interpreted as a
two-dimensional gas with the properties mentioned above. 

The study of the distribution of roots of random polynomials has
been an intensive field of research in recent years
\cite{br}-\cite{fh}. For a Gaussian distribution of
coefficients the problem has been solved. The density \cite{bbl1,bbl2} and all
higher correlation functions \cite{hannay1} have been computed analytically
for arbitrary variances. The gas associated to the zeros of random polynomials
with Gaussian coefficients constitutes therefore a further example in the
restricted class of two-dimensional solvable models (at a certain
temperature). Section 4 is a brief introduction and summary of some results on 
roots of random polynomials.

\section{\hspace*{-0.9cm} .~{\large RANDOM HERMITIAN MATRICES}}

Consider the ensemble of $N \times N$ random hermitian matrices defined in Eq.(\ref{1}).
The basic problem to solve is to find the distribution function of the
associated eigenvalues and eigenvectors. We denote $E_\alpha$,
$\alpha=1,\ldots,N$ the eigenvalues and $\{p_k\}$ a parametrization of the 
eigenvectors of $H$ (the number of parameters $p_k$ depends on the symmetry of
$H$). In the basis where $H$ is diagonal, ${\rm tr} H^2 = 
\sum\limits_{\alpha} E_\alpha^2$. Moreover, the Jacobian of the transformation is
given by \cite{mehta} 
\begin{equation} \label{2}
{\cal J}=\frac{\partial (H_{ij})}{\partial (E_\alpha,p_k)}=\prod_{i<j=1}^N 
|E_i - E_j|^\beta \ .
\end{equation}
The eigenvector's components are absent in this expression because of
the rotational invariance of the ensemble. From these results, we obtain for 
the joint probability density of the eigenvalues (in an
abuse of notation, in this paper we denote all distribution functions with the 
same symbol $P$) 
\begin{equation} \label{3}
P(E_1,\ldots,E_N)=Z^{-1} \prod_{i<j=1}^N |E_i - E_j|^\beta 
\exp\left(- \frac{\beta}{2} \sum_{i=1}^N E_i^2 \right) \ ,
\end{equation}
where $Z$ is a normalization constant.

Interpreting the energies as the (real) positions of some fictitious particles
located at $x_i = E_i$, Eq.(\ref{3}) may be interpreted as the Boltzmann factor
of a classical system, $P(x_1,\ldots,x_N)=Z^{-1} \exp(-\beta V(x_1,\ldots,x_N))$,
with
\begin{equation} \label{4}
V(x_1,\ldots,x_N)=-\sum_{i<j} \log |x_i - x_j| + \frac{1}{2} \sum_i x_i^2 \ .
\end{equation}
The correlations among eigenvalues are therefore described by a
two-dimensional Coulomb system made of unit charges (electrons) restricted to
move over the real axis. In RMT, the temperature takes the discrete values 
$\beta=1/k T=1, 2, 4$.
The harmonic oscillator confining potential is interpreted as the potential
produced by a uniform background of opposite charge density. This is
therefore a one-dimensional one component plasma (1dOCP), with particles'
interactions as in two dimensions.

The correlations introduced by the Jacobian that are responsible for the
Coulomb repulsion among eigenvalues are not a special feature of the particular
ensemble of matrices considered above. In fact, for short distances compared
to the mean level spacing two neighboring eigenvalues of a generic quantum
system repel each other as $|E_+ - E_-|^\beta$. In order to see this, consider
the energy surface of two nearby eigenstates $E_{\pm}$ as a function of a set
of parameters $\{R_1,R_2,\ldots\}$. Two-dimensional
degenerate perturbation theory states that in order to produce a degeneracy,
$(\beta +1)$ parameters should generically be varied. Moreover, near the
degeneracy the contact between the two surfaces has generically the geometry
of a diabolical point (a double cone)
$$
E_\pm \propto \pm \sqrt{\sum_{j=1}^{\beta+1} R_j^2} \ .
$$
It follows from this local geometry that for short distances the probability
$p(s)$ to find two neighboring eigenvalues a distance $s=|E_+ - E_-|$ apart
from each other is proportional to $s^\beta$, as stated above.

From Eqs.(\ref{3}) and (\ref{4}) we see that what RMT does is to extend 
the simple Coulomb interaction which operates at short distances to all 
levels, located at arbitrary distances. It has moreover been shown \cite{bz} 
that the
precise form of the confining potential may modify the density of eigenvalues,
but not the local correlations between eigenvalues (at the scale of several
mean spacings) that are therefore, in this
respect, universal. These universal correlations have been conjectured to
describe the local correlations of eigenvalues of fully chaotic systems 
\cite{bgs,houches1}.

The partition function $Z(\beta)=\int\ldots\int dx_1 \ldots dx_N \exp(-\beta V)$
of the 1dOCP has been computed for any temperature \cite{mehta}
$$
Z(\beta)= \frac{2\pi}{\beta^{N[1+\beta (N-1)/2]/2}} 
\frac{ \prod\limits_{j=1}^N \Gamma (1-\beta j/2)}{\left[ \Gamma
(1+\beta/2)\right]^N} \ .
$$
It as also been shown that the minimum of the potential (\ref{4}) (i.e., the
zero temperature configuration) occurs when the particles are located at the
zeros of the Hermite polynomials of degree $N$.

At the temperatures $\beta=1,2,4$ required by RMT all the correlation functions
of the gas, $R_k (x_1,\ldots,x_k)=(N!/(N-k)!) \int\ldots\int dx_{k+1} \ldots
dx_N P(x_1,\ldots,x_N)$, are known explicitly \cite{mehta}. For example, for
$\beta=2$ and in the large-$N$ limit they take the compact form
\begin{equation} \label{5}
R_k (x_1,\ldots,x_k) = \det \left[\sin [\pi (x_i - x_j)]/[\pi (x_i - x_j)] 
\right]_{i,j=1,\ldots,k}   \ , 
\end{equation}
while the nearest-neighbor spacing distribution function is given by 
\begin{equation} \label{6}
p(s)=\frac{32}{\pi^2} s^2 \exp\left( -\frac{4}{\pi} s^2 \right) \ .
\end{equation}
In fact, this is the $2\times2$ Wigner expression, that happens to be a very
good approximation to the exact one (which takes a more complicated functional
form). For applications of these results and a discussion of their range of
applicability in chaotic and disordered systems see for example  
\cite{houches1,houches2}.

\section{\hspace*{-0.9cm} .~{\large COMPLEX RANDOM MATRICES}}

Here we present an ensemble of random matrices whose (complex) eigenvalues are
described by a fully two-dimensional Coulomb gas, as opposed to eigenvalues of
random hermitian matrices that lie on a line. Another example of a fully
two-dimensional system will be given in next section.

Consider an ensemble $P(H) dH$ of matrices whose matrix elements are
independent complex numbers. For simplicity, we consider again a rotationally
invariant ensemble
\begin{equation} \label{7}
P(H)={\cal N} \exp\left(-\frac{\beta}{2} {\rm tr} (H H^+) \right) \ .
\end{equation} 
Again, starting from this equation the aim is to write down the joint
probability density of the complex eigenvalues $z_i, i=1,\ldots,N$. Similar
steps as in the previous section leads to \cite{ginibre}
\begin{equation} \label{8}
P(z_1,\ldots,z_N)=Z^{-1} \prod_{i<j=1}^N |z_i - z_j|^2 
\exp\left(- \sum_{i=1}^N z_i^2 \right) \ .
\end{equation}
This result is formally identical to the distribution Eq.(\ref{3}) for
$\beta=2$, with the important difference that the eigenvalues now lie in the
complex plane. Its thermodynamical interpretation associates to this
distribution a 2dOCP (a two-dimensional gas of electrons with a uniform
positive background). This analogy has been exploited in Coulomb systems by 
Jancovici \cite{jancovici}.

An explicit formula for the correlations among particles with the distribution
(\ref{8}) can be worked out \cite{ginibre}
\begin{equation} \label{9}
R_k (z_1,\ldots,z_k) = \pi^{-k} \exp\left(-\sum_{i=1}^k |z_i|^2 \right)
                       \det\left[\exp(z_i {\bar z}_j)
                       \right]_{i,j=1,\ldots,k}   \ .
\end{equation}
For $k=1$ and $k=2$ it reads
\begin{equation} \label{10}
\begin{array}{l}
R_1 (z)=\frac{1}{\pi} \exp\left(- |z|^2 \right)
       \sum\limits_{l=0}^{N-1} \displaystyle \frac{|z|^{2 l}}{l!} \ , \\
R_2 (z_1,z_2)=1 - \exp\left(- \pi |z_1 - z_2|^2 \right) \ .
\end{array}
\end{equation}
The density of eigenvalues, that has also been studied by Girko \cite{girko},
tends to a uniform distribution on a disk of radius $|z| \propto \sqrt{N}$, 
with a Gaussian decay outside it. In the expression of $R_2$ the distances 
are measured in units such that the mean spacing is one.

The nearest neighbor spacing distribution of the two-dimensional gas may
be computed from the formula
\begin{equation} \label{11}
\int_0^s p(x) dx = 1-\frac{1}{\kappa}\int\ldots\int_{|z_i|\ge s}
                    d^2 z_2 \ldots d^2 z_N P(z_1=0,z_2,\ldots,z_N) \ .
\end{equation}
The normalization constant $\kappa$ is the probability to find one zero at the
origin, and the notation is $d^2 z = d({\rm Re}z) d({\rm Im}z)$. Evaluating
explicitly the r.h.s. of Eq.(\ref{11}) using Eq.(\ref{8}) one finds 
\cite{ghs}
$$
\int_0^s p(x) dx = 1-\prod_{j=1}^{N-1} \left[ {\rm e}_j (s^2) \exp(-s^2)
\right] \ ,
$$
where ${\rm e}_j (\alpha) = \sum\limits_{k=0}^j \alpha^k /k!$ is the 
truncated exponential. Deriving this expression with respect to $s$ we get
\begin{equation} \label{12}
p(s) = -2 s \prod_{j=1}^{N-1} \left[ \frac{\Gamma (j+1,s^2)}{j!} \right]
            \sum_{k=1}^{N-1} \left[ \frac{k \ \Gamma (k,s^2)}{\Gamma (k+1,s^2)} -
            1 \right] \ ,
\end{equation}
where $\Gamma (n,x) = \int_x^\infty \exp(-t) t^{n-1} dt$ is the
incomplete gamma function. The integral of this function is one, as required.
In order to have also the first moment set to one, one considers the
normalized distribution $p_g (s) = a p(a s)$, with $a=\int_0^\infty s \ p(s) 
ds \approx 1.14293$. The small $s$ expansion of this function
is $p_g (s) = 3.4128 s^3 + {\cal O} (s^5)$.

It is interesting to compare this distribution with the Wigner approximation
in two dimensions, which gives excellent results in one dimension 
(cf Eq.(\ref{6})). This is obtained by solving the approximate equation $p(s)
ds = [1 - \int_0^s p(x) dx ] s^3 ds$. Its solution gives the
normalized distribution 
$$
p_w (s) = 4 \ \Gamma(5/4)^4 \ s^3 \exp\left(-\Gamma(5/4)^4 s^4\right) \ .
$$
For small spacings, it behaves as $p_w (s) = 2.7 s^3 + {\cal O} (s^7)$ 
(see Fig.~1).

\begin{figure}[h]    
\begin{center}
\includegraphics*[scale=0.5,angle=0.0]{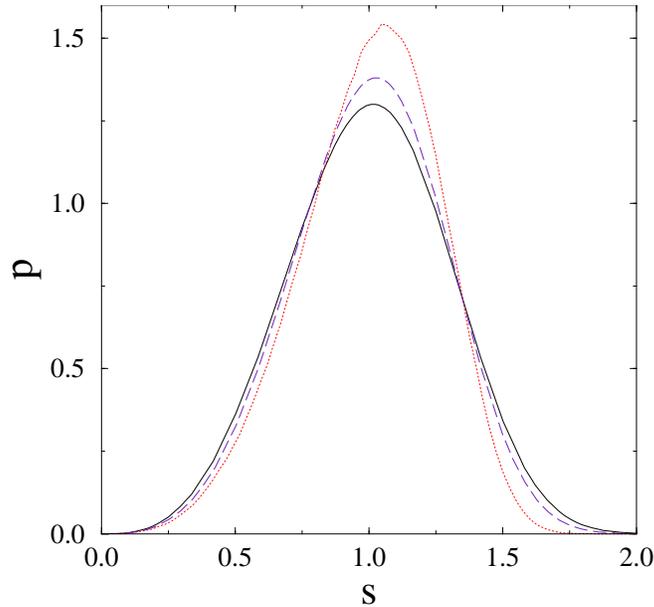}
\caption{The nearest-neighbor spacing distribution for
two-dimensional gases. Full line, $p_g$ (Ginibre); dashed line, $p_w$ (wigner 
approximation); dotted line, $p_{rp}$ (zeros of random polynomials).}

\end{center}
\end{figure}


\section{\hspace*{-0.9cm} .~{\large RANDOM POLYNOMIALS}}

As already pointed out, all ensembles of matrices previously considered are
rotationally invariant. Since there is no privileged direction in Hilbert
space, this induces a very simple form for the probability density
$P(a_0,\ldots,a_N)$ of the coefficients $\{a_j\}$ of an eigenstate $\psi$ of $H$
expressed in an arbitrary basis
\begin{equation} \label{13}
P(a_0,\ldots,a_N) = {\cal C} \ \delta\left( \sum_{j=0}^N |a_j|^2 - 1\right) \ .
\end{equation}
For convenience, we consider now matrices of dimension $(N+1)$;
${\cal C}$ is a normalization constant. We restrict moreover to the
case of complex coefficients (systems without time reversal symmetry,
$\beta=2$).

An interesting way to look at this distribution is the following. Consider the 
Bargmann (or coherent state) representation, in which the eigenstate takes the 
form
\begin{equation} \label{14}
\psi (z) = \sum_{j=0}^N \frac{a_j}{\sqrt{j!}} z^j \ .
\end{equation}
Because we consider here finite dimensional matrices of size $(N+1)$, the
usual series has been truncated. The polynomial (\ref{14}) may be parametrized
by its zeros
$$
 \psi (z) = \frac{a_N}{\sqrt{N!}} \prod_{k=1}^N (z-z_k) \ .
$$
It can easily be shown \cite{kac} that the statistical properties of the
zeros of a polynomial are unchanged if the distribution (\ref{13}) is replaced
by a Gaussian one
\begin{equation} \label{15}
P(a_0,\ldots,a_N) = \prod_{i=0}^N \frac{1}{2 \pi \sigma_i^2} 
                    \exp \left[ -|a_i|^2 /(2 \sigma_i^2) \right] \ .
\end{equation}
The purpose now is, given the probability distribution (\ref{15}) for the
coefficients, to compute the distribution of the zeros of $\psi (z)$ in the
complex plane. Many aspects of this problem for different types of polynomials 
(i.e., having for example some special symmetry, etc) have been understood
in recent years \cite{br}-\cite{fh}.
For the particular case (\ref{15}) of Gaussian independent complex
coefficients, all correlation functions have been explicitly computed 
\cite{bbl1,bbl2,hannay1}. The joint probability density is
easy to write down. Assuming we absorb the factors $1/\sqrt{j!}$ in
Eq.(\ref{14}) into the variance of the coefficients, i.e. we consider
polynomials of the form $\psi (z) = \sum\limits_{j=0}^N a_j \ z^j$, 
with the $\{a_j\}$ distributed according to Eq.(\ref{15}). The joint
probability density is then \cite{bbl2}
\begin{equation} \label{16}
P (z_1,\ldots,z_N) = \\
\frac{N!}{\pi^N \prod\limits_{k=0}^N \sigma_k^2} \
\frac{\prod\limits_{i<j} \left| z_i - z_j \right|^2}{\left(\frac{1}{\sigma_N^2}
+ \frac{\left| z_1+z_2+\ldots \right|^2}{\sigma_{N-1}^2}  + 
\frac{\left|z_1 z_2 + z_1 z_3 + \ldots \right|^2}{\sigma_{N-2}^2}  + \ldots + 
\frac{\left| z_1 z_2 \ldots z_N \right|^2}{\sigma_{0}^2}  \right)^{N+1}} \ .
\end{equation}
Interpreting this distribution as a Boltzmann factor, the numerator
corresponds as in previous sections to a two-dimensional Coulomb interaction
acting among the zeros. This Coulomb interaction is totally generic for zeros 
of polynomials with complex coefficients, since it comes from the Jacobian of
the transformation and does not depend on the particular distribution 
assumed for the coefficients. The difference with Ginibre's gas (the 2dOCP)
lies in the denominator. The gas (\ref{16}) is not confined, like the
eigenvalues of random Gaussian matrices, by a uniform positive background
inducing an harmonic well. The confinement is enforced by a much more
complicated many body interaction between the particles. Contrary to the
numerator, this term depends on the particular form of the distribution of 
the coefficients.

In the particular case Eq.(\ref{14}) where the variances are given by
$\sigma_j^2 = 1/j!$, the density of zeros in the complex plane is
\cite{leboeuf}
$$
R_1 (z) = \frac{1}{\pi} \left\{ 1- g \left( |z|^2 \right) \left[ 1 + N - 
|z|^2 + |z|^2 g \left( |z|^2 \right) \right]  \right\} \ , 
$$
where $g(x)=(x^N /N!)/{\rm e}_N (x)$.
As for the eigenvalues of complex matrices, in the large-$N$ limit this density
is uniform on a disk of radius $\propto \sqrt{N}$. Outside the disk, it tends to
zero much slower than the 2dOCP, since for large radius the tail
of the density decays as  $1/|z|^4$.

A general expression for the two-point correlation function has been derived by
Hannay \cite{hannay1}. For large $N$ and in the uniform-density region it is given by
$$
R_2 (z_1,z_2) =
\frac{\left[ \left( \sinh^2 v + v^2\right) \cosh v - 2 v 
\sinh v \right]}{\sinh^3 v} \ ,
$$
with $v= \pi |z_1 - z_2|^2 /2$ (here again we measure the distances in units
such that the mean spacing is one).

A detailed comparison between the random-polynomial gas and the
2dOCP has been made by Forrester and Honner \cite{fh}. They found that the
first moment $\int d^2 z [R_2 (z,0) - 1] = -1$ in both cases, which means that a perfect screening is operating. However, the second moment differs, 
$\int d^2 z |z^2| [R_2 (z,0) - 1] = -1/\pi$ for the 2dOCP while it vanishes for
the random polynomial gas. In the case of the 2dOCP, this result is known as
the Stillinger-Lovett sum rule. The physical interpretation of the result 
obtained for the zeros of random polynomials is not yet clear. 

The nearest-neighbor spacing distribution for the gas of zeros follows from
Eqs.(\ref{11}) and (\ref{16}). We have not been able to compute the integral.
However, it may be of some practical interest to note that the
numerical curve of $p(s)$ for zeros of random polynomials (in the constant
density region) is well fitted by the expression $p_{rp} (s) = 2.211 s^3 
\exp (-0.389 s^{5.52663})$ (see Fig.~1).

The validity of these results in the study of the structure of eigenstates of
classically chaotic systems has been investigated in \cite{bbl2,ls,prosen}.\\

\section{\hspace*{-0.9cm} .~{\large CONCLUDING REMARKS}}

This brief survey emphasizes the strong connection existing between on the one
hand quantum statistical theories like random matrix theory or random
polynomials and the physics of classical Coulomb systems on the other. The
benefits are mutual. The thermodynamical interpretation serves as a guideline 
in the quantum mechanical studies. In the opposite direction, this connection 
is a powerful tool to compute analytically the correlation functions of the
classical gas, at least at certain temperatures.

Many open problems lie at this interface. One example is the gas
associated to the zeros of random polynomials, whose relevance and physical
interpretation is still unclear from the thermodynamical point of view. This
gas contains, aside a Coulomb repulsion, many body interactions among the
particles. These interactions confine the system (they compensate the strong 
Coulomb repulsion) and also modify the correlations and the local response of
the gas. Nothing is known about its behavior for temperatures different from
$\beta = 2$.

Another example concerns the fact that all the standard ensembles of random 
matrices or polynomials are related to some classical interacting gas with 
particles carrying the same
charge (with eventually a compensating uniform background). It would be
interesting to study ensembles of matrices or polynomials associated to two
(or multi) component plasmas, whose physical behavior is known to be much
richer.\\

\noindent {\bf Acknowledgments}\\

\noindent Le Laboratoire de Physique Th\'eorique et Mod\`eles Statistiques 
est une Unit\'e de recherche de l'Universit\'e de Paris XI associ\'ee au 
CNRS.


\end{document}